\documentstyle[12pt]{article}
\renewcommand{\thefootnote}{\fnsymbol{footnote}}

\newlength{\extraspace}
\newlength{\extraspaces}
\newcommand{\be}{\begin{equation}
\addtolength{\abovedisplayskip}{\extraspaces}
\addtolength{\belowdisplayskip}{\extraspaces}
\addtolength{\abovedisplayshortskip}{\extraspace}
\addtolength{\belowdisplayshortskip}{\extraspace}}
\newcommand{\ee}{\end{equation}}
\newcommand{\ba}{\begin{eqnarray}
\addtolength{\abovedisplayskip}{\extraspaces}
\addtolength{\belowdisplayskip}{\extraspaces}
\addtolength{\abovedisplayshortskip}{\extraspace}
\addtolength{\belowdisplayshortskip}{\extraspace}}
\newcommand{\ea}{\end{eqnarray}}
\newcommand{\newsection}[1]{
\vspace{7mm}
\pagebreak[3]
\addtocounter{section}{1}
\setcounter{equation}{0}
\setcounter{subsection}{0}
{\large {\bf \thesection. #1}}
\nopagebreak
\medskip
\nopagebreak
\hspace{3mm}}
\newcommand{\nonu}{\nonumber \\[.5mm]}
\newcommand{\A}{&\!\!\!}

\begin{document}
%
%\addtolength{\baselineskip}{3.0mm}
%\thispagestyle{empty}

%
\begin{flushright}
SIT-LP-04/02 \\
{\tt gr-qc/0402043} \\
February, 2004
\end{flushright}
\vspace{7mm}

\begin{center}
{\large{\bf{Graded algebraic structure in the canonical formulation 
\\[2mm]
            of $N = 3$ chiral supergravity}}} 
\\[20mm]
{\large Motomu Tsuda}
\footnote{e-mail: tsuda@sit.ac.jp} 
\\[5mm]
Laboratory of Physics, Saitama Institute of Technology \\%[1mm]
Okabe-machi, Saitama 369-0293, Japan \\[20mm]
%
%{\large and}
%
%{\large Takeshi Shirafuji}
%\footnote{e-mail: sirafuji@post.saitama-u.ac.jp} 
%\\
%
%Physics Department, Saitama University \\%[1mm]
%Urawa, Saitama 338-8570, Japan \\[1mm]
%
\begin{abstract}
We focus on $N = 3$ chiral supergravity (SUGRA) 
which is the lowest $N$ theory involving a spin-1/2 field, 
and derive the Ashtekar's canonical formulation of $N = 3$ SUGRA 
starting with the chiral Lagrangian constructed 
by closely following the standard SUGRA. 
The polynomiality of constraints in terms of canonical variables 
and the graded algebraic structure of constraints are discussed 
in the canonical formulation. 
In particular, we show the polynomiality of the {\it rescaled} 
right- and left-handed SUSY constraints by a nonpolynomial factor. 
And also we show the graded algebraic structure of Osp(3/2) 
in the constraint algebra by calculating the Poisson brackets 
of Gauss, SU(2) gauge and right-handed SUSY constraints, 
although the algebra among only those three types 
of constraints does not closed. 
\end{abstract}
\end{center}

%%%%%%%%%%%%%%%%%%%%%%%%%%%%%%%%%%%%%%%%%%%%%%%%%%%%%%%%%%%%
\newpage

\renewcommand{\thefootnote}{\arabic{footnote}}
\setcounter{section}{0}
\setcounter{equation}{0}
\setcounter{footnote}{0}

\newsection{Introduction}

\noindent
Nonperturbative quantum gravity was extensively developed 
in the framework of the Ashtekar's canonical formulation (ACF) 
of general relativity (GR) \cite{AA}, 
which was formulated as an (complexified) SU(2) gauge formulation of GR, 
\footnote{
The canonical formulation of GR based on the real-valued SU(2) connection 
variable was also formulated in \cite{Ba}. 
}
and in the loop quantum gravity (LQG) \cite{JSR,Ro}. 
In those developments, the unification of gravity and gauge fields 
and the fermionic matter contribution 
to the nonperturbative quantum gravity 
have also been discussed; 
gravity and gauge fields (Maxwell or Yang-Mills fields) were discussed 
in \cite{ART}-\cite{CP}, while gravity and spin-1/2 fields were studied 
in \cite{ART,TR}. In \cite{Kr} the Einstein-Maxwell-Dirac theory 
was also considered. 

On the other hand, supersymmetry (SUSY) in both linear realization 
\cite{WZ} and nonlinear realization \cite{VA} is an important notion 
in order to construct an unified theory beyond the standard model. 
From this viewpoint it is useful to investigate the nonperturbative 
aspects of the supergravity (SUGRA) theory as the supersymmetric extension 
of the above works \cite{AA}-\cite{Kr}. 
In fact, the extension of ACF and LQG to SUGRA has been achieved 
by many authors with the following several points 
which have to be discussed; 
namely, 
\begin{enumerate}
\def\labelenumi{(\theenumi)}
\def\theenumi{\alph{enumi}}
\item
the construction of a chiral 
\footnote{In this paper, ``chiral'' means that only right-handed 
(or left-handed) spinor fields are coupled to the self-dual spin 
connection in the kinetic terms of spinor fields. 
}
Lagrangian in first-order form which leads to the ACF of SUGRA, 

\item
the polynomiality of constraints in terms of canonical variables 
in the ACF of SUGRA, 

\item
the graded algebraic structure of constraints 
(in addition to the closure of the contraint algebra) 

\end{enumerate}
and 
\begin{description}
\item{(d)}
quantization and exact solutions of quantum constraints 
(under reality conditions) 

\end{description}
were mainly discussed in the extension. 

The results for (a)-(d) is well-known, in particular, 
up to the extended $N = 2$ chiral SUGRA, 
in which many aspects of ACF and LQG are maintained 
as the supersymmetric extension of \cite{AA}-\cite{CP}. 
Indeed, as for (a), chiral Lagrangians in first-order form were constructed 
for both the $N = 1$ \cite{Ja,CDJ} and $N = 2$ \cite{KS,TS1} theories, 
where a consistency problem arising from the use of 
a complex self-dual spin connection which couples to spinor (spin-3/2) fields 
was solved for $N = 1$ \cite{Ja,TSX,MMT} and for $N = 2$ \cite{TS1}. 

The points of (b)-(d) are the problems 
in the canonical formulation of the chiral SUGRA (i.e., the ACF of SUGRA), 
in which two types of SUSY constraints, right- and left-handed SUSY constraints, 
appear in addition to Gauss-law, U(1) gauge (for $N = 2$), 
vector and Hamiltonian constraints as the result 
of invariances of the chiral Lagrangian. 
Particularly, (c) and (d) show the nonperturbative structures of the chiral SUGRA. 
Indeed, as for (b), in the $N = 1$ theory \cite{Ja} all the constraints 
are written in polynomial form in terms of canonical variables. 
In the $N = 2$ theory \cite{Sa,KS,TS2}, although the left-handed SUSY constraint 
(and also the Hamiltonian constraint) has a nonpolynomial factor 
as in the case of the ACF of the Einstein-Maxwell theory \cite{GP}, 
the rescaled left-handed SUSY constraint by multiplying this factor 
becomes polynomial. 

The simple graded algebraic structure with respect to (c) 
was first pointed out for the $N = 1$ theory through \cite{Fu,UGOP}; 
the SU(2) algebra generated by the Gauss-law constraint 
is graded by means of the right-handed SUSY constraint 
\cite{Fu,UGOP}, and all the constraints were also rewritten 
in a very simple polynomial form \cite{UGOP} 
by using graded connection and momentum variables 
associated with the graded algebra (Lie superalgebra), 
Osp(1/2) (or $G$SU(2)) \cite{PR}. 
The Osp(2/2) (or $G^2$SU(2)) graded algebraic structure \cite{Marcu} 
for the $N = 2$ theory among only the Gauss, U(1) gauge and right-handed SUSY 
constraints was pointed out in \cite{Ez} from the viewpoint of the canonical 
formulation of the BF theory as a toplogical field theory \cite{BTH}, 
and also in \cite{TS2} from the straightforward 
derivation of the canonical formulation of $N = 2$ chiral SUGRA. 
As for exact solutions of quantum constraints in (d), 
two main results of pure gravity, i.e., Wilson loops \cite{JS} 
and the exponential of the Chern-Simons form \cite{KBGP}, 
were discussed both in the $N = 1$ \cite{UGOP} 
and $N = 2$ \cite{Sa,TS2,Ez} theories. 
In addition, for the $N = 1$ theory, based on the irreducible 
representation of Osp(1/2), the spin network state \cite{Pen} 
for SUGRA was constructed in \cite{LiSm}. 

In contrast with the above situation in $N = 1$ and $N = 2$ chiral SUGRA, 
many open questions exist for $N \ge 3$ chiral SUGRA 
except for the construction of the chiral Lagrangian; 
indeed, chiral Lagrangians were constructed 
for $N = 3,\ 4$ theories based on the two-form SUGRA \cite{KN}, 
while for $N = 3,\ 4\ {\rm and}\ 8$ theories 
based on the standard SUGRA \cite{MT}, 
in which Lagrangians do not contain any auxiliary fields as introduced 
in the two-form SUGRA and SUSY transformation parameters 
are not constrained. 
In this paper we focus on $N = 3$ chiral SUGRA 
as the supersymmetric extension of gravity and spin-1/2 fields 
\cite{ART,TR,Kr}, since it is the lowest $N$ theory involving a spin-1/2 field. 
We derive the ACF of $N = 3$ SUGRA by using the chiral Lagrangian constructed 
in \cite{MT}, and we explicitly discuss on the problems (b) and (c), 
i.e., the polynomiality of constraints in terms of canonical variables 
and the graded algebraic structure of constraints for $N = 3$ chiral SUGRA. 
In particular, we show the polynomiality of the {\it rescaled} 
right- and left-handed SUSY constraints by the nonpolynomial factor 
which appears in the ACF of $N = 2$ SUGRA \cite{Sa,KS,TS2}. 
The graded algebraic structure of Osp(3/2) 
(see, for example, Ref.\cite{FSSPr}) 
in the constraint algebra 
is also pointed out by calculating the Poisson brackets 
of the Gauss, SU(2) gauge and right-handed SUSY constraints. 
However, we show that the algebra among only those three types 
of constraints does not closed, 
although the constraint algebra among all the constraints is expected to be closed. 

This paper is organized as follows. 
In Sec.2 we present a globally O(3) invariant 
Lagrangian in $N = 3$ chiral SUGRA which is slightly modified 
from the Lagrangian given in \cite{MT}. 
In Sec.3 a chiral Lagrangian for {\it gauged} $N = 3$ chiral SUGRA 
is introduced by extending the internal, global O(3) 
invariance to local one 
in order to discuss the graded algebraic structure 
of the Gauss, SU(2) gauge and right-handed SUSY constraints. 
The canonical formulation of the gauged $N = 3$ chiral SUGRA 
is derived in Sec.4, and we discuss on the problems (b) and (c) 
in the above arguments. 
We summarize our results in Sec.5.

\newsection{Globally O(3) invariant chiral Lagrangian}

\noindent
In this section we present the Lagrangian of $N = 3$ chiral 
SUGRA \cite{MT}. Corresponding to the spin contents 
$(2, {3 \over 2}, {3 \over 2}, {3 \over 2}, 1, 1, 1, {1 \over 2})$
in the $N = 3$ theory, 
let us denote the fundamental variables as $e^i_{\mu}$ for a (real) tetrad, 
$\psi^{(I)}_{\mu}$ for three (Majorana) Rarita-Schwinger (spin-3/2) fields, 
$A^{(I)}_{\mu}$ for Maxwell fields, 
$\chi$ for a (Majorana) spin-1/2 field 
and $A^{(+)}_{ij \mu}$ for a (complex) self-dual 
spin connection which satisfies 
$(1/2){\epsilon_{ij}} \! ^{kl} A^{(+)}_{kl \mu}$ 
$= i A^{(+)}_{ij \mu}$. 
\footnote{
Greek letters $\mu, \nu, \dots$, are spacetime indices, 
Lattin letters $i, j, \dots$, are local Lorentz indices 
and $(I), (J), \dots (= (1), (2), (3))$, denote O(3) internal indices. 
We take the Minkowski metric $\eta_{ij} = 
{\rm diag}(-1,+1,+1,+1)$ and the totally antisymmetric tensor 
$\epsilon_{ijkl}$ is normalized as $\epsilon_{0123} = +1$. 
We define $\epsilon_{\mu \nu \rho \sigma}$ 
and $\epsilon^{\mu \nu \rho \sigma}$ as tensor 
densities which take values of $+1$ or $-1$.} 
In the first-order formalism, 
the $N = 3$ chiral Lagrangian density in terms of the above 
fundamental variables, which is globally O(3) invariant, 
is written as follows; namely, we have 
\ba
{\cal L}^{(+)}_{N = 3} 
\A = \A - {i \over 2} 
        \epsilon^{\mu \nu \rho \sigma} 
        e^i_{\mu} e^j_{\nu} R^{(+)}_{ij \rho \sigma} 
      - \epsilon^{\mu \nu \rho \sigma} 
        \bar\psi^{(I)}_{R \mu} \gamma_{\rho} 
        D^{(+)}_{\sigma} \psi^{(I)}_{R \nu} 
\nonu
\A \A - {e \over 2} (F^{(-)(I)}{}_{\!\!\! \mu \nu})^2 
      - i e \bar\chi_R \gamma^{\mu} D^{(+)}_{\mu} \chi_R 
\nonu
\A \A + {1 \over{4 \sqrt{2}}} \bar\psi^{(J)}_{\mu} 
        \{ e(F^{(I) \mu \nu} + \hat F^{(I) \mu \nu}) 
        + i \gamma_5 
        (\tilde F^{(I) \mu \nu} + \tilde{\hat F}^{(I) \mu \nu}) \} 
        \psi^{(K)}_{\nu} \epsilon^{(I)(J)(K)} 
\nonu
\A \A + {1 \over 2} e 
        \left( \hat F^{(I)}_{\mu \nu} 
        - {i \over 2} \bar\psi^{(I)}_{\mu} 
        \gamma_{\nu} \chi \right) 
        \bar\psi^{(I)}_{\lambda} 
        S^{\mu \nu} \gamma^{\lambda} \chi 
\nonu
\A \A + {i \over 8} 
        \epsilon^{\mu \nu \rho \sigma} 
        (\bar\psi^{(J)}_{L \mu} \psi^{(K)}_{R \nu}) 
        \bar\psi^{(L)}_{R \rho} \psi^{(M)}_{L \sigma} 
        \epsilon^{(I)(J)(K)} \epsilon^{(I)(L)(M)} 
\nonu
\A \A + {1 \over 8} e 
        (\bar\psi^{(I)}_{\mu} 
        \gamma^\mu \psi^{(I)}_{\nu}) \ 
        \bar\chi \gamma_5 \gamma^\nu \chi, 
\label{LN3}
\ea
where $e$ denotes ${\rm det}(e^i_{\mu})$, 
the covariant derivative $D^{(+)}_\mu$ and 
the curvature ${R^{(+)ij}}_{\mu \nu}$ 
with respect to the $A^{(+)}_{ij \mu}$ are defined by 
\ba
\A \A D^{(+)}_\mu = \partial_\mu + {i \over 2} A^{(+)}_{ij \mu} 
      S^{ij}, \nonu
\A \A {R^{(+)ij}}_{\mu \nu} 
      = 2(\partial_{[\mu} {A^{(+)ij}}_{\nu]} 
      + {A^{(+)i}}_{k [\mu} {A^{(+)kj}}_{\nu]}). 
\ea
In addition, $\epsilon^{(I)(J)(K)}$ is a totally antisymmetric tensor 
and $F^{(I)}_{\mu \nu}$ means the Abelian field strength, 
i.e., $F^{(I)}_{\mu \nu} = 2 \ \partial_{[\mu} A^{(I)}_{\nu]}$. \\
In Eq.(\ref{LN3}), we have used 
$\tilde F^{(I) \mu \nu} = (1/2) \epsilon^{\mu \nu \rho \sigma} 
F^{(I)}_{\rho \sigma}$ and $\hat F_{\mu \nu}$ is defined as 
\be
\hat F^{(I)}_{\mu \nu} 
= F^{(I)}_{\mu \nu} - {1 \over \sqrt{2}} \epsilon^{(I)(J)(K)} 
\bar\psi^{(J)}_\mu \psi^{(K)}_\nu. 
\ee
Following the case of $N = 2$ chiral SUGRA \cite{TS2}, 
we have also used $(F^{(-)(I)}{}_{\!\!\! \mu \nu})^2$ 
as the Maxwell kinetic term in Eq.(\ref{LN3}), where $F^{(-)(I) \mu \nu} 
= (1/2)(F^{(I) \mu \nu} + i e^{-1} \tilde F^{(I) \mu \nu})$. 

Here we explain the role of 
the last two four-fermion contact terms in Eq.(\ref{LN3}) 
with respect to the Rarita-Schwinger 
and spin-1/2 fields \cite{MT}, 
which will be denoted by $\Psi_{\rm 4-fermi}$ 
of Eq.(\ref{p-fermi}) below. 
Those terms are pure imaginary but are necessary to reproduce 
the Lagrangian of the standard $N = 3$ SUGRA \cite{Fr,FSZ} 
in the second-order formalism as follows. 
Indeed, in the first-order formalism, 
the $N = 3$ chiral Lagrangian density (\ref{LN3}) differs 
from that of the standard $N = 3$ SUGRA 
by the following imaginary terms, 
\be
({\cal L}^{(+)}_{N = 3} 
- {\cal L}_{N = 3 {\rm\ standard\ SUGRA}}) \ 
[{\rm first\ order}] 
= \Psi_{\rm kin} + \Psi_{\rm CS-boundary} + \Psi_{\rm 4-fermi} 
\label{diff-1st}
\ee
with $\Psi_{\rm kin}$, $\Psi_{\rm CS-boundary}$ and $\Psi_{\rm 4-fermi}$ 
being defined by 
\ba
\A \A 
\Psi_{\rm kin} = - {i \over 8} \{ \epsilon^{\mu \nu \rho \sigma} 
(T_{\lambda \mu \nu} + i \bar\psi^{(I)}_\mu 
\gamma_\lambda \psi^{(I)}_\nu) T^\lambda{}_{\rho \sigma} 
+ 2 e \bar\chi \gamma_5 \gamma^\nu \chi T^\mu{}_{\mu \nu} \}, 
\label{p-kin} \\
\A \A 
\Psi_{\rm CS-boundary} 
= - {i \over 4} \partial_\mu 
\{ \epsilon^{\mu \nu \rho \sigma} 
(T_{\nu \rho \sigma} + i \bar\psi^{(I)}_\rho 
\gamma_\nu \psi^{(I)}_\sigma) 
- e \bar\chi \gamma_5 \gamma^\mu \chi \nonu
\A \A 
\hspace{2.7cm} + 2 \epsilon^{\mu \nu \rho \sigma} 
A^{(I)}_\nu \partial_\rho A^{(I)}_\sigma \}, 
\label{p-CS} \\
\A \A 
\Psi_{\rm 4-fermi}
= {i \over 8} \epsilon^{\mu \nu \rho \sigma} 
(\bar\psi^{(J)}_{L \mu} \psi^{(K)}_{R \nu}) 
\bar\psi^{(L)}_{R \rho} \psi^{(M)}_{L \sigma} 
\epsilon^{(I)(J)(K)} \epsilon^{(I)(L)(M)} \nonu
\A \A \hspace{1.8cm}
+ {1 \over 8} e 
(\bar\psi^{(I)}_\mu \gamma^\mu \psi^{(I)}_\nu) 
\ \bar\chi \gamma_5 \gamma^\nu \chi, 
\label{p-fermi}
\ea
where the torsion tensor is 
$T^i{}_{\mu \nu} = - 2 D_{[\mu} e^i_{\nu]}$ 
with $D_\mu e^i_\nu = \partial_\mu e^i_\nu 
+ A^i{}_{j \mu} e^j_\nu$. 
The terms in $\Psi_{\rm kin}$ and $\Psi_{\rm CS-boundary}$ appear 
from the chiral gravitational Lagrangian density 
and from the kinetic terms of the Rarita-Shwinger, Maxwell 
and spin-1/2 fields 
in Eq.(\ref{LN3}). 
\footnote{
Note that a boundary term quadratic in 
the Maxwell field $A^{(I)}_{\mu}$ appears in Eq.(\ref{p-CS}) 
since we choose $(F^{(I)(-)}{}_{\!\!\! \mu \nu})^2$ 
as the Maxwell kinetic term in Eq.(\ref{LN3}).} 
The imaginary boundary term $\Psi_{\rm CS-boundary}$ 
of Eq.(\ref{p-CS}) corresponds to a Chern-Simons boundary term 
given in \cite{MMT} as a generating function 
of the canonical transformation. 
However, the four-fermion contact terms 
in $\Psi_{\rm 4-fermi}$ of Eq.(\ref{p-fermi}) do not appear 
in $N = 1$ chiral SUGRA \cite{Ja,MMT}, 
and those are the non-minimal terms required from the invariance 
under first-order SUSY transformations \cite{TS1,MT}. 

In the second-order formalism, 
i.e., when we solve the equation 
$\delta {\cal L}^{(+)}_{N = 3}/\delta A^{(+)}_{ij \mu} = 0$ 
with respect to $A^{(+)}_{ij \mu}$ 
and  we use the obtained solution in Eq.(\ref{diff-1st}), 
e.g., we substitute the solution for the torsion tensor, 
\be
T_{\rho \mu \nu} = -{i \over 2} \bar\psi^{(I)}_\mu 
\gamma_\rho \psi^{(I)}_\nu 
+ {1 \over 4} e \epsilon_{\mu \nu \rho \sigma} 
\bar\chi \gamma_5 \gamma^\sigma \chi, 
\ee
into Eq.(\ref{diff-1st}), 
the $\Psi_{\rm kin}$ of Eq.(\ref{p-kin}) becomes 
\ba
\Psi_{\rm kin} 
= \A \A - {i \over 32} \epsilon^{\mu \nu \rho \sigma} 
  (\bar\psi^{(I)}_\mu \gamma_\lambda \psi^{(I)}_\nu) 
  \bar\psi^{(J)}_\rho \gamma^\lambda 
  \psi^{(J)}_\sigma \mid_{(I) \not= (J)}
\nonu
  \A \A - {1 \over 8} e 
  (\bar\psi^{(I)}_\mu \gamma^\mu \psi^{(I)}_\nu) \ 
  \bar\chi \gamma_5 \gamma^\nu \chi, 
\label{4-Fermi}
\ea
which do not vanish by itself in contrast with 
the $N = 1$ chiral SUGRA. 
On the other hand, the first term in $\Psi_{\rm 4-fermi}$ 
of Eq.(\ref{p-fermi}) can be rewritten as 
\ba
\A \A {i \over 8} \epsilon^{\mu \nu \rho \sigma} 
      (\bar\psi^{(J)}_{L \mu} \psi^{(K)}_{R \nu}) 
      \bar\psi^{(L)}_{R \rho} 
      \psi^{(M)}_{L \sigma} 
      \epsilon^{(I)(J)(K)} \epsilon^{(I)(L)(M)} 
\nonu
%\A \A \ \ \ \ = {i \over 16} \epsilon^{\mu \nu \rho \sigma} 
%      (\bar\psi^{(J)}_{R \mu} \gamma_\lambda \psi^{(L)}_{R \nu}) 
%      \bar\psi^{(K)}_{R \rho} 
%      \gamma^\lambda \psi^{(M)}_{R \sigma} 
%      \epsilon^{(I)(J)(K)} \epsilon^{(I)(L)(M)} 
%\nonu
\A \A \ \ \ \ = {i \over 32} \epsilon^{\mu \nu \rho \sigma} 
  (\bar\psi^{(I)}_\mu \gamma_\lambda \psi^{(I)}_\nu) 
  \bar\psi^{(J)}_\rho \gamma^\lambda 
  \psi^{(J)}_\sigma \mid_{(I) \not= (J)} 
\label{Fierz}
\ea
by using Fierz transformations, 
and so $\Psi_{\rm 4-fermi}$ (i.e., the last two terms in Eq.(\ref{LN3})) 
exactly cancels out the terms of Eq.(\ref{4-Fermi}). 
Therefore, in the second-order formalism, 
the $N = 3$ chiral Lagrangian density (\ref{LN3}) 
is reduced to the Lagrangian density of the $N = 3$ standard SUGRA 
up to imaginary boundary terms as 
\ba
{\cal L}^{(+)}_{N = 3}[{\rm second\ order}] 
= \A \A {\cal L}_{N = 3 {\rm\ standard\ SUGRA}} 
[{\rm second\ order}] \nonu
\A \A 
+ {1 \over 8} \partial_{\mu} 
(\epsilon^{\mu \nu \rho \sigma} 
\bar\psi^{(I)}_\rho \gamma_\nu 
\psi^{(I)}_\sigma 
- i e \bar\chi \gamma_5 \gamma^\mu \chi \nonu
\A \A
- 4i \epsilon^{\mu \nu \rho \sigma} 
A^{(I)}_\nu \partial_\rho A^{(I)}_\sigma). 
\label{LN3S}
\ea

\newsection{Gauging the O(3) invariance}

\noindent
In order to discuss the graded algebraic structure 
in the canonical formulation of $N = 3$ chiral SUGRA 
in the next section, 
let us extend the internal, global O(3) invariance of the chiral Lagrangian 
density (\ref{LN3}) to local one \cite{FrDa}. 
This method, i.e., gauging the O(3) invariance, is the same 
as the case of $N = 2$ SUGRA \cite{FrDa} except for the introduction 
of the non-Abelian field strength. 
Indeed, after introducing a minimal coupling of $\psi^{(I)}_{R \mu}$ 
with $A^{(I)}_{\mu}$, 
it requires to replace the Abelian field 
strength $F^{(I)}_{\mu \nu}$ with the non-Abelian one, 
\be
F'^{(I)}_{\mu \nu} = F^{(I)}_{\mu \nu} 
+ \lambda \epsilon^{(I)(J)(K)} A^{(J)}_\mu A^{(K)}_\nu 
\label{non-Ab}
\ee
with the gauge coupling constant $\lambda$. 
Furthermore the minimal coupling automatically 
requires a spin-3/2 mass-like term and a cosmological term 
in order to ensure the SUSY invariance of the Lagrangian; 
these three terms added to Eq.(\ref{LN3}) are then written as 
\ba
{\cal L}_{{\rm cosm}} 
= \A \A {\lambda \over 2} \epsilon^{\mu \nu \rho \sigma} 
        \bar\psi^{(I)}_\mu 
        \gamma_5 \gamma_\rho \psi^{(K)}_\nu A^{(J)}_\sigma 
        \epsilon^{(I)(J)(K)} 
\nonu
\A \A - \sqrt{2} i \kappa^{-1} \lambda e 
        (\bar\psi^{(I)}_\mu 
        S^{\mu \nu} \psi^{(I)}_\nu ) 
\nonu
\A \A - \Lambda \kappa^{-2} e, 
%+ 6 \kappa^{-4} \lambda^2 e, 
\label{Lcosm}
\ea
where the cosmological constant $\Lambda$ is related 
to $\lambda$ as $\Lambda = - 6 \kappa^{-2} \lambda^2$. 
Note that the first term of Eq.(\ref{Lcosm}) is 
comparable with the kinetic term of $\psi^{(I)}_{R \mu}$ 
in Eq.(\ref{LN3}), because 
\be
{\lambda \over 2} \epsilon^{\mu \nu \rho \sigma} 
        \bar\psi^{(I)}_\mu 
        \gamma_5 \gamma_\rho \psi^{(K)}_\nu A^{(J)}_\sigma 
        \epsilon^{(I)(J)(K)} 
= - \lambda \epsilon^{\mu \nu \rho \sigma} 
        \bar\psi^{(I)}_{R \mu} 
        \gamma_\rho \psi^{(K)}_{R \nu} A^{(J)}_\sigma 
        \epsilon^{(I)(J)(K)}. 
\ee
We denote the {\it gauged} $N = 3$ chiral Lagrangian density 
as the sum of Eqs.(\ref{LN3}) and (\ref{Lcosm}), 
in which $F^{(I)}_{\mu \nu}$ is replaced by $F'^{(I)}_{\mu \nu}$ 
of Eq.(\ref{non-Ab}); namely, we define 
\be
{\cal L}^{(+){\rm gauged}}_{N = 3} 
= {\cal L}^{(+)}_{N = 3}[F^{(I)}_{\mu \nu} \rightarrow F'^{(I)}_{\mu \nu}] 
+ {\cal L}_{{\rm cosm}}. 
\label{L+}
\ee

Let us give two comments on the above gauged chiral Lagrangian. 
First, from the discussion in Sec.2, 
it is obvious that in the second-order formalism 
the gauged $N = 3$ chiral Lagrangian density (\ref{L+}) 
is reduced to the gauged Lagrangian density of the $N = 3$ standard SUGRA 
\cite{FrDa} up to imaginary boundary terms as 
\ba
\A \A {\cal L}^{(+){\rm gauged}}_{N = 3}[{\rm second\ order}] 
= {\cal L}^{{\rm gauged}}_{N = 3 {\rm\ standard\ SUGRA}} 
[{\rm second\ order}] 
\nonu
\A \A \hspace{1cm} + {1 \over 8} \partial_{\mu} 
(\epsilon^{\mu \nu \rho \sigma} 
\bar\psi^I_\rho \gamma_\nu 
\psi^I_\sigma 
- i e \ \bar\chi \gamma_5 \gamma^\mu \chi) 
\nonu
\A \A \hspace{1cm} - {i \over 2} \partial_\mu 
\left\{ \epsilon^{\mu \nu \rho \sigma} 
\left( A^{(I)}_\nu \partial_{\rho} A^{(I)}_\sigma 
+ {\lambda \over 3} \epsilon^{(I)(J)(K)} 
A^{(I)}_\nu A^{(J)}_\rho A^{(K)}_\sigma \right) 
\right\}. 
\label{LN3SS}
\ea
Secondly, because of Eq.(\ref{LN3SS}), 
the ${\cal L}^{(+){\rm gauged}}_{N = 3}$ of Eq.(\ref{L+}) in the second-order 
formalism is invariant under the SUSY transformation 
of the standard gauged $N = 3$ SUGRA given in \cite{FrDa} by 
\ba
\delta e^i_\mu 
= \A \A i \ \bar\alpha^{(I)} 
  \gamma^i \psi^{(I)}_\mu, 
\nonu
\delta A^{(I)}_\mu 
= \A \A \sqrt{2} \ \epsilon^{(I)(J)(K)} 
  \bar\alpha^{(J)} \psi^{(K)}_\mu + i \bar\alpha^{(I)} \gamma_\mu \chi, 
\nonu
\delta \psi^{(I)}_\mu 
= \A \A 2 \{ D_\mu[A(e, \psi^{(I)})] \alpha^{(I)} 
  + \lambda \epsilon^{(I)(J)(K)} A^{(J)}_\mu \alpha^{(K)} \} 
- {i \over 8} (\bar\chi \gamma_5 \gamma^\lambda \chi) 
\gamma_5 \gamma_\lambda \gamma_\mu \alpha^{(I)} 
\nonu
\A \A - {1 \over \sqrt{2}} \epsilon^{(I)(J)(K)} 
\bar F'^{(J)}_{\rho \sigma} S^{\rho \sigma} \gamma_\mu \alpha^{(K)} 
\nonu
\A \A + {1 \over {2 \sqrt{2}}} \epsilon^{(I)(J)(K)} 
\{ (\bar\psi^{(J)}_\mu \gamma_\nu \chi) \gamma^\nu \alpha^{(K)} 
- (\bar\psi^{(J)}_\mu \gamma_5 \gamma_\nu \chi) 
\gamma_5 \gamma^\nu \alpha^{(K)} \} 
\nonu
\A \A - \sqrt{2} i \ \lambda \gamma_\mu \alpha^{(I)}, 
\nonu
\delta \chi = \A \A -i \bar F'^{(I)}_{\mu \nu} S^{\mu \nu} \alpha^{(I)}, 
\label{SUSY-N3}
\ea
where $A_{ij \mu}(e, \psi^{(I)})$ 
in $\delta \psi^{(I)}_\mu$ is defined as the sum 
of the Ricci rotation coefficients $A_{ij \mu}(e)$ 
and $K_{ij \mu}$ which is expressed as 
\be
K_{\mu \nu \rho} = {i \over 4} 
(\bar\psi^{(I)}_\mu \gamma_\rho \psi^{(I)}_\nu 
+ \bar\psi^{(I)}_\mu \gamma_\nu \psi^{(I)}_\rho 
- \bar\psi^{(I)}_\nu \gamma_\mu \psi^{(I)}_\rho) 
+ {1 \over 8} e \epsilon_{\mu \nu \rho \sigma} 
\bar\chi \gamma_5 \gamma^\sigma \chi. 
\ee
On the other hand, in the first-order formalism, 
the SUSY invariance of ${\cal L}^{(+){\rm gauged}}_{N = 3}$ 
may be realized by introducing the right- and left-handed 
SUSY transformations as in the case of 
$N = 1$ chiral SUGRA \cite{Ja,TS3}.

\newsection{Canonical formulation 
of $N = 3$ chiral SUGRA}

\noindent
In this section, we derive the canonical formulation 
of $N = 3$ chiral SUGRA (the ACF of $N = 3$ SUGRA) 
by means of the (3+1) decomposition of spacetime, 
starting with the gauged $N = 3$ chiral Lagrangian density (\ref{L+}). 
The gauge condition for the tetrad $e^i_\mu$ 
in the (3+1) decomposition of spacetime which we shall follow 
is the same as that of \cite{TS2}. 
\footnote{As for the indices of the canonical formulation, 
Latin letters $a, b, \cdots$ are used as 
the spatial part of the spacetime indices $\mu, \nu, \cdots$, 
and capital letters {\it I, J,} $\cdots$ are used 
as the spatial part of the local Lorentz indices {\it i, j,} $\cdots$. 
Two-component spinor indices $A, B, \dots$ 
and $A', B', \dots$ are also used. 
As for the conventions of the two-component spinor formulation 
and the other several conventions in the canonical formulation, 
we shall follow those of \cite{TS2}. 
}
Namely, we assume that the topology of spacetime 
$M$ is $\Sigma \times R$ for some three-manifold $\Sigma$ 
so that a time coordinate function $t$ is defined on $M$. 
Then the time component of the tetrad can be defined as 
%
%\footnote{Latin letters $a, b, \cdots$ are 
%the spatial part of the spacetime indices 
%$\mu, \nu, \cdots$, and capital letters 
%{\it I, J,} $\cdots$ denote the spatial part 
%of the local Lorentz indices {\it i, j,} $\cdots$.}
%
%
\be
e^i_t = N n^i + N^a e^i_a. 
\ee
Here $n^i$ is the timelike unit vector orthogonal to 
$e_{ia}$, i.e., $n^i e_{ia} = 0$ and $n^i n_i = - 1$, 
while $N$ and $N^a$ denote the lapse function 
and the shift vector, respectively. 
Furthermore, we give a restriction on the tetrad with 
the choice $n_i = (- 1, 0, 0, 0)$ in order to simplify 
the Legendre transform of Eq.(\ref{L+}). 
Once this choice is made, $e_{Ia}$ becomes tangent to 
the constant $t$ surfaces $\Sigma$ and $e_{0a} = 0$. 
Therefore we change the notation $e_{Ia}$ 
to $E_{Ia}$ below. We also take the spatial restriction 
of the totally antisymmetric tensor density 
$\epsilon^{\mu \nu \rho \sigma}$ 
as $\epsilon^{abc} = \epsilon{_t}^{abc}$, 
while $\epsilon^{IJK} = \epsilon{_0}^{IJK}$. 

From the (3+1) decomposition of Eq.(\ref{L+}) 
under the above gauge condition of the tetrad, 
the kinetic terms in the canonical formulation 
which define canonical variables are obtained as 
%
%\footnote{
%For convenience of calculation we use the two-component 
%spinor notation in the canonical formulation. 
%Two-component spinor indices $A, B, \dots$, 
%and $A', B', \dots$, are raised and lowered 
%with antisymmetric spinors $\epsilon^{AB}, \epsilon_{AB}$, 
%and their conjugates $\epsilon^{A'B'}, \epsilon_{A'B'}$, 
%such that $\psi^A = \epsilon^{AB} \psi_B$ and 
%$\varphi_B = \varphi^A \epsilon_{AB}$. 
%The Infeld-van der Waerden symbol $\sigma_{iAA'}$ are 
%taken in this paper to be $(\sigma_0, \sigma_I) 
%:= (-i/\sqrt{2}) (I, \tau_I)$ with $\tau_I$ being 
%the Pauli matrices. We also define the symbol 
%$\sigma_{IA}{}^B$ (which is called the $SU(2)$ soldering 
%form in \cite{AA}) by using $n^{AA'} = n^i \sigma{_i}^{AA'}$ 
%as $\sigma_{IA}{}^B := - \sqrt{2} i \sigma_{IAA'} n^{BA'} 
%= (i/\sqrt{2})(\tau_I)_{AB}$.}
%
\be
{\cal L}^{(+){\rm gauged}}_{N = 3}[{\rm kinetic\ terms}] 
= \tilde E_I^a \dot {\cal A}^I{}_a 
- \tilde \pi^{(I)}{}_A{}^a \dot \psi^{(I)A}{}_a 
+ {}^+ \tilde \pi^{(I)a} \dot A^{(I)}_a 
- \tilde \pi_A \dot \chi^A, 
\label{Lkin}
\ee
where ${\cal A}^I{}_a := - 2 A^{(+)}{}_0{}^I{}_a$ 
and the momentum variables ($\tilde \pi^{(I)}{}_A{}^a$, 
${}^+ \tilde \pi^{(I)a}$, $\tilde \pi_A$) are defined by 
%
%\footnote{
%The derivative for fermionic variables is treated 
%as the left derivative unless stated otherwise.}
%
\ba
\tilde \pi^{(I)}{}_A{}^a 
= \A \A {{\delta {\cal L}^{(+)}} 
\over {\delta \dot \psi^{(I)A}{}_a}} 
= - \sqrt{2} i \ \epsilon^{abc} E^I_c \ 
\bar\psi^{(I)A'}{}_b \sigma_{IAA'}, 
\\ 
{}^+ \tilde \pi^{(I)a} 
= \A \A {{\delta {\cal L}^{(+)}} 
\over {\delta \dot A^{(I)}_a}} 
= \tilde \pi^{(I)a} + i \ \tilde B'^{(I)a} 
\\
\tilde \pi_A 
= \A \A {{\delta^L {\cal L}^{(+)}} 
\over {\delta \dot \chi^A}} 
= \sqrt{2} \ E \bar\chi^{A'} n_{AA'} 
\ea
with 
\ba
\tilde \pi^{(I)a} 
= \A \A {e \over {2 N^2}} q^{ab} 
\{ \ 2 \ (F'^{(I)}_{tb} - N^d F'^{(I)}_{db}) 
\nonu
\A \A - \sqrt{2} \ (\bar\psi^{(J)}_t \psi^{(K)}_b 
- N^d \bar\psi^{(J)}_d \psi^{(K)}_b) 
\epsilon^{(I)(J)(K)} \} 
\nonu
\A \A - {i \over {2 \sqrt{2}}} \ \epsilon^{abc} 
\bar\psi^{(J)}_b \gamma_5 \psi^{(K)}_c 
\epsilon^{(I)(J)(K)} 
\nonu
\A \A - {{i e} \over {2 N^2}} q^{ab} 
(E^I_b \bar\psi^{(I)}_t \gamma_I \chi 
- N^d E^I_b \overline \psi^{(I)}_d \gamma_I \chi 
- N \bar\psi^{(I)}_b \gamma_0 \chi) 
\nonu
\A \A - {1 \over 2} \epsilon^{abc} E^I_c 
\bar\psi^{(I)}_b \gamma_5 \gamma_I \chi, 
\label{pi}
\\
\tilde B'^{(I)a} 
= \A \A {1 \over 2} \ \epsilon^{abc} F'^{(I)}_{bc}. 
\ea
We have used the Majorana spinors $\psi^{(I)}_{\mu}$ in Eq.(\ref{pi}) 
for simplicity. 

Furthermore, we obtain constraints which reflect the invariance 
of the gauged $N = 3$ chiral Lagrangian density (\ref{L+}) 
from the variation of the Lagrangian with respect to Lagrange multipliers. 
In this paper, let us explicitly show the Gauss, SU(2) gauge, 
right-handed SUSY and left-handed SUSY constraints 
in terms of the canonical variables. 
Varying Eq.(\ref{L+}) by Lagrange multipliers 
$\Lambda^I_t$, $A^{(I)}_t$, $\psi^{(I)A}{}_t$ and $\rho^{(I)A}{}_t$, 
in which $\Lambda^I_t$ and $\rho{^{(I)A}}_t$ are defined by 
\be
\Lambda^I_t = - 2 A^{(+)}_0{}^I{}_t, \ \ \ \ \ 
\rho^{(I)A}{}_t = E^{-1} \bar\psi^{(I)}_{A't} n^{AA'}, 
\ee
yields those four types of constraints as 
%
%\footnote{
%We note that the ${}^+ \tilde \pi^{(I)a}$ appears 
%in Eqs.(\ref{SU2}), (\ref{RSUSY}) and (\ref{LSUSY}). 
%If we use $(F^{(I)}_{\mu \nu})^2$ as the Maxwell kinetic 
%term in Eq.(\ref{LN3}), the $\tilde \pi^{(I)a}$ 
%(and not ${}^+ \tilde \pi^{(I)a}$) will appear 
%in Eq.(\ref{SU2}), and it is not possible to rewrite 
%the canonical constraints in terms of 
%the graded variables.}
%
\ba
{\cal G}_I 
= \A \A {{\delta {\cal L}^{(+)}} \over {\delta \Lambda^I_t}} 
= {\cal D}_a \tilde E_I^a 
- {i \over \sqrt{2}} \ \tilde \pi^{(I)}{}_A{}^a 
\sigma{_I}{^A}_B \psi{^{(I)B}}_a 
- {i \over \sqrt{2}} \ \tilde \pi_A 
\sigma{_I}{^A}_B \chi^B = 0, 
\label{Gauss} 
\\
g^{(I)} 
= \A \A {{\delta {\cal L}^{(+)}} \over {\delta A^{(I)}_t}} 
= \partial_a {}^+ \tilde \pi^{(I)a} 
+ \lambda \epsilon^{(I)(J)(K)} A^{(J)}_a 
{}^+ \tilde \pi^{(K)a} 
+ \lambda \ \psi^{(J)A}{}_a 
\tilde \pi^{(K)}{}_A{}^a \epsilon^{(I)(J)(K)} 
\nonu
= \A \A 0, 
\label{SU2} 
\\
{}^R {\cal S}^{(I)}_A 
= \A \A {{\delta^L {\cal L}^{(+)}} 
\over {\delta \psi^{(I)A}{}_t}} 
\nonu
= \A \A {\cal D}_a \tilde \pi^{(I)}{}_A{}^a + {1 \over \sqrt{2}} 
\ {}^+ \tilde \pi^{(K)a} \ \psi{^{(J)B}}_a 
\epsilon_{AB} \epsilon^{(I)(J)(K)} 
\nonu
\A \A + \lambda \ (2i \tilde E_I^a \sigma^I{}_{AB} 
\psi^{(I)B}{}_a - \tilde \pi^{(J)}{}_A{}^a 
A^{(K)}_a \epsilon^{(I)(J)(K)}) 
\nonu
\A \A + {1 \over 2} E^{-2} \epsilon_{ade} 
{(\sigma^I \sigma^J)}_A{}^B \tilde E_I^d \tilde E_J^e \tilde \pi_B 
\Phi^a = 0, 
%\nonu
%\A \A \times \left\{ \epsilon^{abc} 
%\left( F'^{(I)}_{bc} + {1 \over \sqrt{2}} \ \epsilon_{CD} 
%\psi^{(J)C}{}_b \psi^{(K)D}{}_c \epsilon^{(I)(J)(K)} \right) 
%+ i \ {}^+ \tilde \pi^{(I)a} + i \tilde \pi^{(I)}{}_C{}^a \chi^C \right\} 
%\nonu
%= \A \A 0, 
\label{RSUSY} 
\\
{}^L {\cal S}^{(I)}_A 
= \A \A {{\delta {\cal L}^{(+)}} \over {\delta \rho^{(I)A}{}_t}} 
\nonu
= \A \A - \sqrt{2} 
\ \tilde E_I^a \tilde E_J^b {(\sigma^I \sigma^J)}_A{}^B 
\Bigg\{ \ 2({\cal D}_{[a} \psi^{(I)C}{}_{b]} 
+ \lambda \ A^{(J)}_{[a} \psi^{(K)C}{}_{b]} 
\epsilon^{(I)(J)(K)}) \epsilon_{BC} 
\nonu
\A \A \left. + {i \over \sqrt{2}} 
\lambda \ \epsilon_{abc} \tilde \pi^{(I)}{}_B{}^c 
- {i \over 2} \epsilon_{abc} {}^+ \tilde \pi^{(I)c} 
\chi^C \epsilon_{BC} \right\} 
\nonu
\A \A + {i \over 2} E^{-2} \epsilon_{def} \epsilon_{agh} 
\epsilon^{(I)(J)(K)} {(\sigma^I \sigma^J \sigma^K \sigma^L)}_A{}^B 
\tilde E_I^e \tilde E_J^f \tilde E_K^g \tilde E_L^h 
\tilde \pi^{(J)}{}_B{}^d 
\Phi^a = 0, 
%\nonu
%\A \A \times \left\{ \epsilon^{abc} \left( F'^{(K)}_{bc} 
%+ {1 \over \sqrt{2}} \ \epsilon_{CD} \psi^{(L)C}{}_b \psi^{(M)D}{}_c 
%\epsilon^{(K)(L)(M)} \right) 
%+ i \ {}^+ \tilde \pi^{(K)a} + i \tilde \pi^{(K)}{}_C{}^a \chi^C 
%\right\} 
%\nonu
%= \A \A 0, 
\label{LSUSY}
\ea
where the $\Phi^a$ in Eqs.(\ref{RSUSY}) and (\ref{LSUSY}) 
is a quantity expressed by the canonical variables 
in polynomial form as 
\ba
\Phi^a 
= \A \A \epsilon^{abc} \left( F'^{(K)}_{bc} 
+ {1 \over \sqrt{2}} \ \epsilon_{CD} \psi^{(L)C}{}_b \psi^{(M)D}{}_c 
\epsilon^{(K)(L)(M)} \right) 
\nonu
\A \A + i {}^+ \tilde \pi^{(K)a} + i \tilde \pi^{(K)}{}_C{}^a \chi^C, 
\ea
and the covariant derivatives on $\Sigma$ are defined as 
\ba
\A \A {\cal D}_a \tilde E_I^a 
= \partial_a \tilde E_I^a + i \epsilon_{IJK} 
{\cal A}^J{}_a \tilde E^{Ka}, 
\nonu
\A \A {\cal D}_a \tilde \pi^{(I)}{}_A{}^a 
= \partial_a \tilde \pi^{(I)}{}_A{}^a 
- {i \over \sqrt{2}} {\cal A}_A{}^B{}_a 
\tilde \pi^{(I)}{}_B{}^a. 
\ea
Obviously, the Gauss and SU(2) gauge constraints 
of Eqs.(\ref{Gauss}) and (\ref{SU2}) are polynomial 
with respect to the canonical variables, 
while both the right- and left-handed SUSY constraints 
of Eqs.(\ref{RSUSY}) and (\ref{LSUSY}) 
are not polynomial because of the factor $E^{-2}$. 
But the {\it rescaled} right- and left-handed SUSY 
constraints by the nonpolynomial factor, 
i.e., $E^2 \ {}^R {\cal S}^{(I)}_A$ and $E^2 \ {}^L {\cal S}^{(I)}_A$ 
become polynomial because of the relation, 
$E^2 = (1/6) \epsilon_{abc} \epsilon^{IJK} 
\tilde E_I^a \tilde E_J^b \tilde E_K^c$. 

In order to discuss the graded algebraic structure 
in the canonical formulation of $N = 3$ chiral SUGRA, 
we calculate the Poisson brackets of 
the Gauss, SU(2) gauge and right-handed 
SUSY constraints of Eqs. from (\ref{Gauss}) to (\ref{RSUSY}) 
by using the non-vanishing Poisson brackets 
%
%\footnote{The Poisson brackets are defined 
%for canonical variables $(q^i, \tilde p_i)$ 
%by using the right and left derivatives as 
%$\{ F, G \} := \int d^3 z [(\delta^R F/\delta q^i(z)) 
%(\delta^L G/\delta \tilde p_i(z)) 
%- (-1)^{\mid i \mid} (\delta^R F/\delta \tilde p_i(z)) 
%(\delta^L G/\delta q^i(z))]$ 
%with $\mid i \mid = 0$ for an even (commuting) $q^i$ 
%while $\mid i \mid = 1$ for an odd (anticommuting) $q^i$.}
%
among the canonical variables, 
\ba
\A \A \{ {\cal A}^I{}_a(x), \tilde E_J{}^b(y) \} 
= \delta^I_J \delta_a^b \delta^3(x - y), 
\nonu
\A \A \{ \psi^{(I)A}{}_a(x), 
\tilde \pi^{(J)}{}_B{}^b(y) \} 
= - \delta^{(I)(J)} \delta_B^A \delta_a^b \delta^3(x - y), 
\nonu
\A \A \{ A^{(I)}_a(x), {}^+ \tilde \pi^{(J)b}(y) \} 
= \delta^{(I)(J)} \delta_a^b \delta^3(x - y), 
\nonu
\A \A \{ \chi^A(x), \tilde \pi_B(y) \} 
= - \delta_B^A \delta^3(x - y). 
\ea
In fact, when we define the smeared functions, 
\ba
\A \A {\cal G}_I [\Lambda^I] 
= \int_\Sigma d^3 x \ \Lambda^I {\cal G}_I, 
\nonu
\A \A g^{(I)} [a^{(I)}] 
= \int_\Sigma d^3 x \ a^{(I)} \ g^{(I)}, 
\nonu
\A \A (E^2 \ {}^R {\cal S}^{(I)}_A) [\xi^{(I)A}] 
= \int_\Sigma d^3 x \ \xi^{(I)A} \ (E^2 \ {}^R {\cal S}^{(I)}_A) 
\ea
for convenience of the calculation, 
the Poisson brackets of 
${\cal G}_I$, $g^{(I)}$ and ${}^R {\cal S}^{(I)}_A$ 
are obtained as 
\ba
\A \A 
\{ {\cal G}_I [\Lambda^I], \ {\cal G}_J [\Gamma^J] \} 
= {\cal G}_I [\Lambda'^I], 
\label{GG}
\\
\A \A 
\{ {\cal G}_I [\Lambda^I], \ g^{(I)} [a^{(I)}] \} = 0, 
\\
\A \A 
\{ g^{(I)} [a^{(I)}], \ g^{(J)} [b^{(J)}] \} 
= \lambda g^{(I)} [a'^{(I)}], 
\label{gg}
\\
\A \A 
\{ {\cal G}_I [\Lambda^I], 
\ (E^2 \ {}^R {\cal S}^{(I)}_A) [\xi^{(I)A}] \} 
= (E^2 \ {}^R {\cal S}^{(I)}_A) [\xi'^{(I)A}], 
\\
\A \A 
\{ g^{(I)} [a^{(I)}], \ (E^2 \ {}^R {\cal S}^{(J)}_A) [\xi^{(J)A}] \} 
= \lambda \ (E^2 \ {}^R {\cal S}^{(I)}_A) [\xi''^{(I)A}], 
\\
\A \A 
\{ (E^2 \ {}^R {\cal S}^{(I)}_A) [\xi^{(I)A}], 
\ (E^2 \ {}^R {\cal S}^{(J)}_B) [\eta^{(J)B}] \} 
= \lambda E^4 \ {\cal G}_I [\Lambda''^I] + E^4 \ g^{(I)} [a''^{(I)}] 
\nonu
\A \A \hspace{4cm} 
+ E^2 \ {}^R {\cal S}^{(I)}_A [\eta'^{(I)A}] 
+ E^2 \ {}^L {\cal S}^{(I)}_A [\eta''^{(I)A}], 
\label{RR}
\ea
where the smeared function, 
\be
(E^2 \ {}^L {\cal S}^{(I)}_A) [\xi^{(I)A}] 
= \int_\Sigma d^3 x \ \xi^{(I)A} \ (E^2 \ {}^L {\cal S}^{(I)}_A) 
\ee
has also been used in Eq.(\ref{RR}). 
In Eqs. from (\ref{GG}) to (\ref{RR}), $\Lambda'^I$, $\Lambda''^I$, 
$\xi'^{(I)A}$, $\xi''^{(I)A}$, $a'^{(I)}$ and $a''^{(I)}$, are defined 
as the field-independent parameters by 
\ba
\A \A 
\Lambda'^I = i \ \epsilon^{IJK} \Lambda_J \Gamma_K, 
\nonu
\A \A 
\Lambda''^I = 2i \ \xi^{(I)A} \eta^{(J)B}
\sigma^I{}_{AB} \delta^{(I)(J)}, 
\nonu
\A \A 
\xi'^{(I)A} = {i \over \sqrt{2}} \ 
\Lambda^I \xi^{(I)B} \sigma_{IB}{}^A, 
\nonu
\A \A 
\xi''^{(I)A} = \epsilon^{(I)(J)(K)} a^{(J)} \ \xi^{(K)A}, 
\nonu
\A \A 
a'^{(I)} = \epsilon^{(I)(J)(K)} a^{(J)} b^{(K)}, 
\nonu
\A \A 
a''^{(I)} = {1 \over \sqrt{2}} \ \epsilon^{(I)(J)(K)} 
\xi^{(J)A} \eta^{(K)B} \epsilon_{AB}, 
\ea
and also $\eta'^{(I)A}$ and $\eta''^{(I)A}$ 
are defined as the field-dependent parameters by 
\ba
\A \A 
\eta'^{(I)A} = - {i \over 2} 
(\xi^{(I)A} \eta^{(J)B} + \xi^{(J)B} \eta^{(I)A}) 
\epsilon^{abc} (\sigma^I \sigma^J)_B{}^C 
\tilde E_I^a \tilde E_J^b \tilde \pi^{(J)}{}_C{}^c, 
\nonu
\A \A 
\eta''^{(I)A} = {1 \over 2} \epsilon^{(I)(J)(K)} 
\xi^{(J)B} \eta^{(K)C} \epsilon_{BC} \tilde \pi^A. 
\ea
Except for the last two terms in the r.h.s. of Eq.(\ref{RR}), 
the resultant Poisson bracket (\ref{GG})-(\ref{RR}) 
shows that the SU(2) $\times$ SU(2) algebra of Eqs. from (\ref{GG}) 
to (\ref{gg}) is graded by means of ${}^R {\cal S}^{(I)}_A$, 
i.e., the algebra of the Gauss, SU(2) gauge 
and right-handed SUSY constraints includes the graded algebra 
(Lie superalgebra), Osp(3/2) \cite{FSSPr}. 
This is expressed in terms of the generators 
$(J_I, J^{(I)}; J^{(I)}_A)$ which correspond to those three types 
of constraints as 
\ba
\A \A [J_I, J_J] = i \epsilon_{IJ}{}^K J_K, 
\nonu
\A \A [J^{(I)}, J^{(J)}] = \lambda \epsilon^{(I)(J)(K)} J^{(K)}, 
\nonu
\A \A [J_I, J^{(I)}_A] 
= {i \over \sqrt{2}} \sigma_{IA}{}^B J^{(I)}_B, 
\nonu
\A \A [J^{(I)}, J^{(J)}_A] 
= \lambda \epsilon^{(I)(J)(K)} J^{(K)}_A, 
\nonu
\A \A [J^{(I)}_A, J^{(J)}_B] 
= 2i \lambda \delta^{(I)(J)} \sigma^I{}_{AB} J_I 
+ {1 \over \sqrt{2}} \epsilon^{(I)(J)(K)} \epsilon_{AB} J^{(K)}, 
\nonu
\A \A 
[J_I, J^{(I)}] = 0, 
\ea
where the structure constants are determined from Eqs.(\ref{GG})-(\ref{RR}). 
However, in contrast with the case of the $N = 1$ and $N = 2$ theories, 
the algebra among only the Gauss, SU(2) and right-handed SUSY 
constraints does not closed, 
in particular, by the last term in Eq.(\ref{RR}) 
which is proportional to the left-handed SUSY constraint 
${}^L {\cal S}^{(I)}_A$, 
although the algebra among all the constraints 
which appear in the canonical formulation is expected to be closed.

\newsection{Conclusions}

\noindent
In this paper we have derived the canonical formulation 
of $N = 3$ chiral SUGRA (the ACF of $N = 3$ SUGRA), 
starting with the gauged $N = 3$ chiral Lagrangian density (\ref{L+}), 
in which the Maxwell kinetic term is modified 
as $(F'^{(-)(I)}{}_{\!\!\! \mu \nu})^2$. 
We have shown the explicit form 
of the Gauss, SU(2) gauge, right-handed SUSY and left-handed SUSY 
constraints in terms of the canonical variables, 
and we have also discussed 
that both the right- and left-handed SUSY constraints 
are not polynomial because of the nonpolynomial factor $E^{-2}$, 
but the rescaled constraints by this factor become polynomial. 
In additon, by calculating of the Poisson brackets 
among the Gauss, SU(2) gauge and right-handed SUSY constraints 
following the case of the canonical formulation 
of $N = 1$ and $N = 2$ chiral SUGRA, 
we have shown the graded algebraic structure 
of Osp(3/2) \cite{FSSPr} in the constraint algebra. 
However, in contrast with the case of the $N = 1$ and $N = 2$ theories, 
the algebra among only those three types 
of constraints does not closed, in particular, 
by the term which is proportional to the left-handed SUSY constraint, 
although the algebra among all the constraints 
is expected to be closed as a whole. 

We are now trying to canonically quantize the theory 
and to derive exact solutions of quantum constraints, 
e.g., based on the introduction of graded variables associated 
with the Osp(3/2) algebra. 

\vspace{2cm}

\noindent
{\large{\bf{Acknowledgments}}} 

\noindent
I am grateful to Professor T. Shirafuji for useful discussions 
and encouragements at Physics Department of Saitama University. 
Also I would like to thank Professor K. Shima for useful discussions 
and encouragements.

%%%%%%%%%%%%%%%%%%%%%%%%%%%%%%%%%%%%%%%%%%%%%%%%%%%%%%%%%%%%%%%%%%%%%
%%%%%%%%%%%%%%%%%%%%%%%%%%%%%%%%%%%%%%%%%%%%%%%%%%%%%%%%%%%%%%%%%%%%%

%%%%%%%%%%%%%%%%%%%%%%%%%%%%%%%%%%%%%%%%%%%%%%%%%%%%%%%%%%%%%%%%%%%%%
%%%%%%%%%%%%%%%%%%%%%%%%%%%%%%%%%%%%%%%%%%%%%%%%%%%%%%%%%%%%%%%%%%%%%


\newpage
\begin{thebibliography}{100}

\bibitem{AA}
%A. Ashtekar, 
%Phys. Rev. Lett. {\bf 57}, 2244 (1986); 
%Phys. Rev. D {\bf 36}, 1587 (1987). 
Ashtekar A 1986 {\it Phys. Rev. Lett.} {\bf 57} 2244 \\
Ashtekar A 1987 {\it Phys. Rev.} D {\bf 36} 1587 

\bibitem{Ba}
%J. F. Barbero G., 
%Phys. Rev. D {\bf 51}, 5507 (1995). 
Barbero J F 1995 {\it Phys. Rev.} D {\bf 51} 5507 

\bibitem{JSR}
%T. Jacobson and L. Smolin, 
%Nucl. Phys. B {\bf 299}, 295 (1988); \\
%C. Rovelli and L. Smolin, 
%Phys. Rev. Lett. {\bf 61}, 1155 (1988); \\
%C. Rovelli and L. Smolin, 
%Nucl. Phys. B {\bf 331}, 80 (1990). 
Jacobson T and Smolin L 1988 {\it Nucl. Phys.} B {\bf 299} 295 \\
Rovelli C and Smolin L 1988 {\it Phys. Rev. Lett.} {\bf 61} 1155 \\
Rovelli C and Smolin L 1990 {\it Nucl. Phys.} B {\bf 331} 80 

\bibitem{Ro}
%C. Rovelli, {\it Loop quantum gravity}, gr-qc/9710008 
%(review written for the electronic journal 
%{\it Living Reviews} No 1998-1)
Rovelli C 1997 Loop quantum gravity {\it Preprint} gr-qc/9710008 
(review written for the electronic journal {\it Living Reviews} No 1998-1)

\bibitem{ART}
%A. Ashtekar, J. D. Romano and R. S. Tate, 
%Phys. Rev. D {\bf 40}, 2572 (1989). 
Ashtekar A, Romano J D and Tate R S 1989 
{\it Phys. Rev.} D {\bf 40} 2572 

\bibitem{GP}
%R. Gambini and J. Pullin, 
%Phys. Rev. D {\bf 47}, R5214 (1993). 
Gambini R and Pullin J 1993 
{\it Phys. Rev.} D {\bf 47} R5214 

\bibitem{CP}
%S. Chakraborty and P. Peld\'an, 
%Phys. Rev. Lett. {\bf 73}, 1195 (1994). 
Chakraborty S and Peld\'an P 1994 
{\it Phys. Rev. Lett.} {\bf 73} 1195 

\bibitem{TR}
%H. A. Morales-T\'ecotl and C. Rovelli, 
%Phys. Rev. Lett. {\bf 72}, 3642 (1994); 
%Nucl. Phys. B {\bf 451}, 325 (1995). 
Morales-T\'ecotl H A and Rovelli C 1994 
{\it Phys. Rev. Lett.} {\bf 72} 3642 \\
Morales-T\'ecotl H A and Rovelli C 1995 
{\it Nucl. Phys.} B {\bf 451} 325 

\bibitem{Kr}
%K. V. Krasnov, 
%Phys. Rev. D {\bf 53}, 1874 (1996). 
Krasnov K V 1996 {\it Phys. Rev.} D {\bf 53} 1874 

\bibitem{WZ}
%J. Wess and B. Zumino, 
%Phys. Lett. {\bf B49}, 52 (1974). 
Wess J and Zumino B 1974 {\it Phys. Lett.} B {\bf 49} 52 

\bibitem{VA}
%D.V. Volkov and V.P. Akulov,  
%JETP Lett. {\bf 16}, 438 (1972); 
%Phys. Lett. {\bf B46}, 109 (1973). 
Volkov D V and Akulov V P  1972 
{\it JETP Lett.} {\bf 16} 438 \\
Volkov D V and Akulov V P  1973 
{\it Phys. Lett.} {\bf B46} 109 

\bibitem{Ja}
%T. Jacobson, 
%Class. Quantum Grav. {\bf 5}, 923 (1988). 
Jacobson T 1988 {\it Class. Quantum Grav.} {\bf 5} 923 

\bibitem{CDJ}
%R. Capovilla, J. Dell, T. Jacobson and L. Mason, 
%Class. Quantum Grav. {\bf 8}, 41 (1991). 
Capovilla R, Dell J, Jacobson T and Mason L 1991 
{\it Class. Quantum Grav.} {\bf 8}, 41 

\bibitem{Sa}
%T. Sano, hep-th/9211103. 
Sano T {\it Preprint} hep-th/9211103 

\bibitem{KS}
%H. Kunitomo and T. Sano, 
%Prog. Theor. Phys. Suppl. {\bf 114}, 31 (1993); 
%Int. J. Mod. Phys. {\bf D} 1, 559 (1993). 
Kunitomo H and Sano T 1993 
{\it Prog. Theor. Phys. Suppl.} {\bf 114} 31 \\
Kunitomo H and Sano T 1993 
{\it Int. J. Mod. Phys.} D {\bf 1} 559 

\bibitem{TS1}
%M. Tsuda and T. Shirafuji, 
%Class. Quantum Grav. {\bf 16}, 69 (1999). 
Tsuda M and Shirafuji T 1999 
{\it Class. Quantum Grav.} {\bf 16} 69 

\bibitem{TSX}
%M. Tsuda, T. Shirafuji and H. Xie, 
%Class. Quantum Grav. {\bf 12}, 3067 (1995). 
Tsuda M, Shirafuji T and Xie H 1995 
{\it Class. Quantum Grav.} {\bf 12} 3067 

\bibitem{MMT}
%A. Mac\'\i as, 
%Class. Quantum Grav. {\bf 13}, 3163 (1996); \\
%E. W. Mielke, A. Mac\'\i as and H. A. Morales-T\'ecotl, 
%Phys. Lett. A {\bf 215}, 14 (1996). 
Mac\'\i as A 1996 {\it Class. Quantum Grav.} {\bf 13} 3163 \\
Mielke E W, Mac\'\i as A and Morales-T\'ecotl H A 1996 
{\it Phys. Lett.} {\bf A215} 14 (1996) 

\bibitem{TS2}
%M. Tsuda and T. Shirafuji, 
%Phys. Rev. D {\bf 62}, 064020 (2000). 
Tsuda M and Shirafuji T 2000 
{\it Phys. Rev.} D {\bf 62} 064020 

\bibitem{Fu}
%G. F\"ul\"op, 
%Class. Quantum Grav. {\bf 11}, 1 (1994). 
F\"ul\"op G 1994 {\it Class. Quantum Grav.} {\bf 11} 1 

\bibitem{UGOP}
%D. Armand-Ugon, R. Gambini, O. Obreg\'on and J. Pullin, 
%Nucl. Phys. B {\bf 460}, 615 (1996). 
Armand-Ugon D, Gambini R, Obreg\'on O and Pullin J 1996 
{\it Nucl. Phys.} B {\bf 460} 615 

\bibitem{PR}
%A. Pais and V. Rittenberg, 
%J. Math. Phys. {\bf 16}, 2062 (1975). 
Pais A and Rittenberg V 1975 
{\it J. Math. Phys.} {\bf 16} 2062 

\bibitem{Marcu}
%M. Marcu, 
%J. Math. Phys. {\bf 21}, 1277 (1980). 
Marcu M 1980 {\it J. Math. Phys.} {\bf 21} 1277 

\bibitem{Ez}
%K. Ezawa, 
%Prog. Theor. Phys. {\bf 114}, 31 (1996). 
Ezawa K 1996 {\it Prog. Theor. Phys.} {\bf 114} 31 

\bibitem{BTH}
%M. Blau and G. Thompson, 
%Phys. Lett. {\bf B228}, 64 (1989); 
%Ann. Phys. (N.Y.) {\bf 205}, 130 (1991); \\
%G. Horowitz, 
%Commun. Math. Phys. {\bf 125}, 417 (1989). 
Blau M and Thompson G 1989 
{\it Phys. Lett.} {\bf B228} 64 \\
Blau M and Thompson G 1991 
{\it Ann. Phys. (N.Y.)} {\bf 205} 130 \\
Horowitz G 1989 
{\it Commun. Math. Phys.} {\bf 125} 417 

\bibitem{JS}
%T. Jacobson and L. Smolin, 
%Nucl. Phys. B {\bf 299}, 295 (1988). 
Jacobson T and Smolin L 1988 
{\it Nucl. Phys.} B {\bf 299} 295 

\bibitem{KBGP}
%H. Kodama, 
%Phys. Rev. D {\bf 42}, 2548 (1990); \\
%B. Br\"ugmann, R. Gambini and J. Pullin, 
%Nucl. Phys. B {\bf 385}, 587 (1992). 
Kodama H 1990 {\it Phys. Rev.} D {\bf 42} 2548 \\
Br\"ugmann B, Gambini R and Pullin J 1992 
{\it Nucl. Phys.} B {\bf 385} 587 

\bibitem{Pen}
%R. Penrose, 
%in {\it Quantum Theory and Beyond}, 
%edited by T. Bastin (Cambridge Univ. Press, 
%Cambridge, 1992); 
%C. Rovelli and L. Smolin, 
%Phys. Rev. D {\bf 52}, 5743 (1995). 
Penrose R 1992 
in {\it Quantum Theory and Beyond} 
edited by Bastin T (Cambridge, Cambridge Univ. Press) \\
Rovelli C and Smolin L 1995 
{\it Phys. Rev.} D {\bf 52} 5743 

\bibitem{LiSm}
%Y. Ling and L. Smolin, 
%Phys. Rev. D {\bf 61}, 044008 (2000). 
Ling Y and Smolin L 2000 
{\it Phys. Rev.} D {\bf 61} 044008 

\bibitem{KN}
%T. Kadoyoshi and S. Nojiri, 
%Mod. Phys. Lett. A {\bf 12}, 1165 (1997). 
Kadoyoshi T and Nojiri S 1997 
{\it Mod. Phys. Lett.} A {\bf 12} 1165 

\bibitem{MT}
%M. Tsuda, 
%Phys. Rev. D {\bf 63}, 104021 (2001). 
Tsuda M 2001 {\it Phys. Rev.} D {\bf 63} 104021 

\bibitem{FSSPr}
%L. Frappat, P. Sorba and A. Sciarrino, 
%{\it Dictionary on Lie superalgebras}, 
%hep-th/9607161; \\
%A. V. Proeyen, {\it Tools for supersymmetry}, 
%hep-th/9910030. 
Frappat L, Sorba P and Sciarrino A 1996 
Dictionary on Lie superalgebras {\it Preprint} hep-th/9607161 \\
Proeyen A V 1999 Tools for supersymmetry {\it Preprint} hep-th/9910030 

\bibitem{Fr}
%D. Z. Freedman, 
%Phys. Rev. Lett. {\bf 17}, 105 (1977). 
Freedman D Z 1977 {\it Phys. Rev. Lett.} {\bf 17} 105 

\bibitem{FSZ}
%S. Ferrara, J. Scherk and B. Zumino, 
%Phys. Lett. B {\bf 66}, 35 (1977). 
Ferrara S, Scherk J and Zumino B 1977 
{\it Phys. Lett.} B {\bf 66} 35 

\bibitem{FrDa} 
%D. Z. Freedman and A. Das, 
%Nucl. Phys. B {\bf 120}, 221 (1977). 
Freedman D Z and Das A 1977 
{\it Nucl. Phys.} B {\bf 120} 221 

\bibitem{TS3} 
%M. Tsuda and T. Shirafuji, 
%Phys. Rev. D {\bf 54}, 2960 (1996). 
Tsuda M and Shirafuji T 1996 
{\it Phys. Rev.} D {\bf 54} 2960 





\end{thebibliography}
\end{document}